\documentclass[12pt]{article}
\pdfoutput=1

\usepackage{draft} 
\usepackage{hyperref}
\usepackage{graphicx,color,subfig}
\usepackage{cite}
\usepackage{mciteplus}
\usepackage{skak}
\DeclareFontFamily{OT1}{pzc}{}
\DeclareFontShape{OT1}{pzc}{m}{it}{<-> s * [1.10] pzcmi7t}{}
\DeclareMathAlphabet{\mathpzc}{OT1}{pzc}{m}{it}

\def\cst{${\cal CS}^2$}

\numberwithin{equation}{section}

\def\e{{\varepsilon}}

 \def\p{\partial}
 \def\bz{{\bar z}}
 
\def\0{{(0)}}
\def\1{{(1)}}
\def\2{{(2)}}

\def\<{\langle }
\def\>{\rangle }

\def\eads${H$_3$}

\def\be#1\ee{\begin{align}#1\end{align}}

\begin{document}

\unitlength = .8mm

\begin{titlepage}

\begin{center}

\hfill \\
\hfill \\
\vskip 1cm

\title{Flat Space Amplitudes\\
and Conformal Symmetry of the Celestial Sphere}


\author{Sabrina Pasterski,$^1$ Shu-Heng Shao,$^2$ and Andrew Strominger$^1$}

\address{
$^1$Center for the Fundamental Laws of Nature, Harvard University,\\
Cambridge, MA 02138, USA
\\
$^2$School of Natural Sciences, Institute for Advanced Study, \\Princeton, NJ 08540, USA
}

\end{center}

\vspace{2.0cm}

\begin{abstract}

The four-dimensional (4D) Lorentz group $SL(2,\mathbb{C})$ acts as the two-dimensional (2D) global conformal group on the celestial sphere at infinity where asymptotic 4D scattering states are specified. 
Consequent similarities of 4D flat space amplitudes and 2D correlators on the conformal sphere are obscured by the fact that the former are usually expressed in terms of asymptotic wavefunctions which transform simply under spacetime translations rather than the Lorentz $SL(2,\mathbb{C})$.
In this paper we construct on-shell massive scalar wavefunctions in 4D Minkowski space that transform as $SL(2,\mathbb{C})$ conformal primaries.  Scattering amplitudes of these wavefunctions are $SL(2,\mathbb{C})$ covariant by construction.  For certain mass relations, we show explicitly that their three-point amplitude reduces to the known unique form of  a 2D CFT primary three-point function and compute the coefficient. The computation proceeds naturally via Witten-like diagrams on a hyperbolic slicing of Minkowski space and has a holographic flavor.

\end{abstract}

\vfill

\end{titlepage}

\eject
\tableofcontents

\section{Introduction}

Quantum field theory (QFT) scattering amplitudes in four-dimensional (4D) Minkowski space are typically expressed in terms of asymptotic plane wave solutions to the free wave equation. Translation invariance is manifest because the plane waves simply acquire phases which cancel due to momentum conservation. The $SL(2,\mathbb{C})$ Lorentz invariance is more subtle as plane waves transform non-trivially into one another. 
In this paper we find a basis of $SL(2,\mathbb{C})$ primary solutions to the massive scalar wave equation  and recast certain  4D scattering amplitudes in a manifestly $SL(2,\mathbb{C})$ covariant form. This form is very familiar from the the study of two-dimensional (2D) conformal field theory (CFT), in which $SL(2,\mathbb{C})$ acts as the global conformal group. The appearance of the 2D conformal group is no coincidence, since the Lorentz group  acts as the global conformal group on the celestial sphere, denoted by \cst, at null infinity where the asymptotic states are specified. 

Studies of the $SL(2,\mathbb{C})$ properties of scattering amplitudes date back to Dirac \cite{Dirac:1936fq}, but a new reason has arisen for interest in the topic. When gravity is coupled, it was conjectured in \cite{deBoer:2003vf,Banks:2003vp,Barnich:2009se,Barnich:2011ct} that the $SL(2,\mathbb{C})$ is enhanced to the full infinite-dimensional  
local conformal (or Virasoro)  group.  This conjecture was recently proven \cite{Kapec:2014opa,Kapec:2016jld,Cheung:2016iub} to follow at tree level\footnote{The subleading soft theorem has a one-loop exact anomaly \cite{He:2014bga,Bianchi:2014gla,Bern:2014oka,Bern:2014vva}  whose effects remain to be understood  but are recently discussed in \cite{Hawking:2016sgy,loop}.} from the new subleading soft graviton theorem of \cite{Cachazo:2014fwa}.\footnote{One may hope that ultimately 4D quantum gravity scattering amplitudes are found to have a dual holographic representation as some exotic 2D CFT on \cst, but at present there are no proposals for such a construction.} While the full Virasoro appears only when gravity is coupled, the scattering amplitudes of any QFT that  can be 
coupled to gravity are constrained to be `Virasoro-ready'.\footnote{This generalizes the well known constraint that any QFT that can be coupled to gravity must have a local conserved stress tensor.} This suggests that they should resemble  a subset of 2D CFT correlators, likely involving complex and continuous conformal dimensions. 
Indeed it has already been observed \cite{Strominger:2013lka,He:2015zea}  that soft photon amplitudes take the form of a 2D current algebra. Here we seek to understand the 2D description of 4D scattering amplitudes away from the soft limit.

This paper considers massive scalar three-point functions, and recasts them in the standard form of 2D CFT correlators on the celestial sphere \cst. To summarize the result, let $X^\mu~(\mu=0,1,2,3)$ be the flat coordinates on Minkowski space.   A natural coordinate on \cst\ where $X^\mu X_\mu=0$ is  
\begin{equation}\label{sdf} w={X^1+iX^2 \over X^0+X^3}\,.\end{equation}
Lorentz transformations act as the global conformal group on \cst\ 
\be w \to {aw+b\over cw+d}\,.\ee
Here the complex parameters $a,b,c,d$ obey $ad-bc=1$. We will construct a three-parameter family of solutions labelled by a point $w$ on \cst\ and an $SL(2,\mathbb{C})$  weight $\Delta$  (rather than the three components of spatial momenta $\vec p$ which label plane waves) which transform as conformal primaries. We will find below they are naturally displayed in a hyperbolic slicing of Minkowski space, in harmony with the form of flat space holography advocated in \cite{deBoer:2003vf}. 
$SL(2,\mathbb{C})$ then  implies that the  4D tree amplitude takes the form  
\be  \mathcal{ \tilde A} (w_i,\bar w_i)= {C\over |w_1-w_2|^{\Delta_1+\Delta_2-\Delta_3}|w_2-w_3|^{\Delta_2+\Delta_3-\Delta_1}|w_3-w_1|^{\Delta_3+\Delta_1-\Delta_2}}\,,\ee
where the `OPE coefficients'  $C$ depends on the masses, conformal weights, and cubic coupling of the three asymptotic scalars. An integral formula for $C$ involving Witten-like diagrams on the hyperbolic slices is derived. 
 In general this integral may not be computable in closed form, but it simplifies in the near-extremal case when the incoming particle is only slightly heavier than the sum of the outgoing particles and is explicitly given below. 
 
 Other tractable examples would be of great interest. In particular, the beautiful structure of 
 ${\cal N}=4$ amplitudes suggest they may take a particularly nice 2D form rewritten as correlators on \cst.  In \cite{Cheung:2016iub} one contribution to such amplitudes (from the interior of the forward lightcone) was expressed as a Witten diagram, but to obtain the full amplitude additional harder-to-compute contributions (from outside the lightcone) must be added in. This remains an outstanding open problem. 
 
 The utility of hyperbolic slicing was already noticed in a context with some similarity to the present one by Ashtekar and Romano in \cite{Ashtekar:1991vb}. de Boer and Solodukhin initiated a program to understand flat space holography in terms of AdS holography via hyperbolic slicing in \cite{deBoer:2003vf}. 
Soft theorems and aspects of scattering were hyberbolically studied  in \cite{Campiglia:2015qka,Campiglia:2015kxa,Campiglia:2015lxa,Cheung:2016iub}. 
 In the context of the recent revival of the conformal bootstrap program, the linear realization of the conformal symmetry in the embedding  Minkowski space has been used to simplify computations of, for example,  conformal blocks and  propagators  in AdS \cite{Cornalba:2009ax,Weinberg:2010fx,Costa:2011mg,Costa:2011dw,SimmonsDuffin:2012uy}. 

The outline of the paper is as follows. In Section \ref{sec:wavefunction} we define and construct the massive and massless scalar wavefunctions that are conformal primaries of the Lorentz group $SL(2,\mathbb{C})$. Our construction, equation (\ref{integral}) below, is a convolution of plane waves with the bulk-to-boundary propagator on the hyperbolic slice $H_3$, and is evaluated in terms of Bessel functions. We also present an integral transform  that takes a massive scalar scattering amplitude into an $SL(2,\mathbb{C})$ covariant correlation function. In Section \ref{sec:3pt} we compute the  three-point amplitude  of massive conformal primary wavefunctions in the near-extremal limit.  The main result is equation (\ref{main}). In Appendix  \ref{sec:KG} we compute the Klein-Gordon inner product of these primary wavefunctions for a fixed mass.

\section{Conformal Primary Wavefunctions}\label{sec:wavefunction}

In this section we construct  scalar wavefunctions that are conformal primaries of the Lorentz group $SL(2,\mathbb{C})$. 
A scalar \textit{conformal primary wavefunction} $\phi_{\Delta,m}(X^\mu; w,\bar w)$ of mass $m$ and conformal dimension $\Delta$ is defined by the following two properties:
\begin{enumerate}
\item It is a solution to the  Klein-Gordon equation of mass $m$,\footnote{We will use the $(-,+,+,+)$ convention for the signature in $\mathbb{R}^{1,3}$.}
\begin{align}
\left({\partial \over \partial X^\nu}{\partial \over \partial X_\nu} -m^2\right) \phi_{\Delta,m}(X^\mu;w,\bar w)=0\,.
\end{align}
\item It transforms covariantly as a conformal (quasi-)primary operator in two dimensions under an $SL(2,\mathbb{C})$ Lorentz transformation,
\begin{align}\label{highestweight}
\phi_{\Delta,m}\left( \Lambda^\mu_{~\nu} X^\nu ; {aw+b\over cw+d} , {\bar a \bar w+ \bar b \over \bar c\bar w+ \bar d}\right)
= |cw+d|^{2\Delta}\phi_{\Delta,m}\left( X^\mu ; w,\bar w\right)\,,
\end{align}
where $a,b,c,d\in \mathbb{C}$ with $ad-bc=1$ and $\Lambda^\mu_{~\nu}$ is its associated $SL(2,\mathbb{C})$ group element in the four-dimensional representation.\footnote{There is no canonical way to embed the celestial sphere into the lightcone in Minkowski space.  It follows that there is also no canonical way to associate a  M\"obius action on $w$ with an $SL(2,\mathbb{C})$ element $\Lambda^\mu_{~\nu}$ in the four-dimensional representation. In fact, any two $\Lambda^\mu_{~\nu}$'s that differ by an $SL(2,\mathbb{C})$ conjugation are equally good for our definition. 
  Below  we will make a choice of the map $\Lambda^\mu_{~\nu}(a,b,c,d)$ by fixing a reference frame for  $\hat p^\mu$ in \eqref{map}. More explicitly, $\Lambda^\mu_{~\nu}$ is the $SL(2,\mathbb{C})$ transformation matrix acting on $\hat p^\mu$ given by plugging \eqref{sl2c} into \eqref{map}.}

\end{enumerate}
Note that, in contrast to the situation in AdS/CFT, the mass $m$ and the conformal dimension $\Delta$ are not  related.

\subsection{Integral Representation}

Conformal primary wavefunctions for a massive scalar field can be constructed from the Fourier transform of the bulk-to-boundary propagator in three-dimensional hyperbolic space $H_3$. Let $(y,z,\bar z)$ be the  Poincar\'e coordinates of the $H_3$ with metric,
\begin{align}\label{ads3metric}
ds_{H_3}^2 = {dy^2 +dzd\bar z\over y^2}\,.
\end{align}
Here $0<y<\infty$ and  $y=0$ is the  boundary of the $H_3$.  This geometry  has an $SL(2,\mathbb{C})$ isometry that acts as
\begin{align}\label{sl2c}
&z\to z' = {(az+b)(\bar c\bar z +\bar d )+a\bar cy^2 \over|cz+d|^2+|c|^2y^2 } \,,\notag\\
&\bar z\to \bar z' = {(\bar a\bar z+\bar b)(c z + d )+\bar a cy^2 \over|cz+d|^2+| c|^2y^2 } \,,\notag\\
&y\to y' = {y\over |cz+d|^2+|c|^2y^2}\,,
\end{align}
with $a,b,c,d\in \mathbb{C}$ and $ad-bc =1$.  The $H_3$  can be embedded into $\mathbb{R}^{1,3}$ as either one of the two branches ($\hat p^0>0$ or $\hat p^0<0$) of the unit hyperboloid
\begin{align}\label{hyperboloid}
 -(\hat p^0)^2 + (\hat p^1)^2+(\hat p^2)^2+(\hat p^3)^2=-1\,.
\end{align}
 More explicitly, we can choose this embedding map $\hat p^\mu:$ $H_3$$\to \mathbb{R}^{1,3}$  for the upper hyperboloid, corresponding to an outgoing particle,  to be
\begin{align}\label{map}
\hat p^\mu(y,z,\bar z) = \left({  1+y^2+|z|^2 \over 2y}  ,  {\text{Re} (z)\over y}, {\text{Im} (z)\over y}, {1-y^2-|z|^2\over 2y}    \right)\,.
\end{align}
This implies the useful relation
\begin{equation}z={\hat p^1+i\hat p^2 \over\hat p^0+\hat p^3}\,.\end{equation}

Let $G_\Delta(y,z,\bar z; w,\bar w)$ be the scalar bulk-to-boundary propagator of conformal dimension $\Delta$ in $H_3$ \cite{Witten:1998qj}, 
\begin{align}
G_\Delta (y, z,\bar z; w,\bar w) =  \left({ y\over y^2  +  |z-w|^2}\right)^\Delta\,.
\end{align}
This transforms covariantly  under the $SL(2,\mathbb{C})$ transformation \eqref{sl2c},
\begin{align}\label{covariance}
&G_\Delta(y',z',\bar z'; w' ,\bar w') = |cw+d|^{2\Delta} G_\Delta(y,z,\bar z; w,\bar w)\,,
\end{align}
where $w' = (aw+b)/(cw+d)$ and $\bar w'  = (\bar a\bar w +\bar b)/(\bar c \bar w +\bar d)$.

The conformal primary wavefunction for a massive scalar is 
\begin{align}\label{integral}
\boxed{\,
\phi^\pm_{\Delta,m} (X^\mu; w, \bar w) 
= \int_0^\infty   {dy \over y^3} \int dzd\bar z  \, 
G_\Delta (y ,z,\bar z;w, \bar w)  \,
\exp\Big[   { \pm i m \,\hat p^\mu(y,z,\bar z)\, X_\mu}   \Big]\,,\,}
\end{align}
where we pick the  minus (plus) sign for an incoming (outgoing) particle.   In the next subsection we will see that potentially divergent integrals can be regulated in an $SL(2,\mathbb{C})$ covariant manner by complexifying the mass $m$ and (\ref{integral}) is expressed in terms of Bessel functions.

It is trivial to check that \eqref{integral} is indeed a conformal primary wavefunction. First, it satisfies the massive Klein-Gordon equation because each plane wave factor $e^{i m \hat p\cdot X}$ does. Second, it is a conformal quasi-primary (in the sense of \eqref{highestweight}) because of the $SL(2,\mathbb{C})$ covariance \eqref{covariance} of the  bulk-to-boundary propagator in $H_3$.    
Our definition and formula for the conformal primary wavefunction \eqref{integral} can be readily  generalized to Minkowski space of any dimension $\mathbb{R}^{1,d+1}$ and it would transform covariantly under the Euclidean $d$-dimensional conformal group $SO(1,d+1)$.

The $H_3$  is embedded, via the map \eqref{map}, into the hyperboloid in the momentum space, rather than position space and the boundary point $w,\bar w$ might seem to live at the boundary of momentum space rather than Minkowski space. However, these spaces are canonically identified. The trajectory of a free massive particle is 
\be X^\mu(s)=\hat p^\mu s +X_0^\mu\, .\ee
At late times, $s \to \infty,~-X^2\to \infty$ and 
\be {X^\mu \over \sqrt{-X^2}}\to {\hat p^\mu }\,.\ee
That is, massive particles asymptote to a fixed position on the hyperbolic slices of Minkowski determined by their four-momenta. Hence $(w,\bar w)$ can be interpreted as a boundary coordinate of the late-time asymptotic $H_3$ slice.

Although we are far from constructing any example of such, the authors of \cite{deBoer:2003vf} speculate on a boundary 2D CFT (of some exotic variety) on \cst\ a subset of whose correlation functions are the 4D bulk Minkowski scattering amplitudes. Every bulk field would be dual  to a continuum of operators labelled by their conformal weights. In this putative 2D CFT, the scalar bulk field mode \eqref{integral} would be holographically dual to a local scalar boundary operator  ${\cal O}_{\Delta}(w,\bar w)$ of dimension $\Delta$. 

The $SL(2,\mathbb{C})$ covariance of the conformal primary wavefunction  implies the  $SL(2,\mathbb{C})$ covariance of any scattering amplitudes constructed from them. 
Let $p^\mu_j$ be the on-shell momenta of $n$ massive scalars of masses $m_j$ ($j=1,\cdots, n$).    
Given any Lorentz invariant $n$-point momentum-space scattering amplitude $\mathcal{A}(p_j^\mu)$ of these massive scalars (including the momentum conservation delta function $\delta^{(4)}(\sum_{j=1}^n  p_j^\mu)$), the conformal primary amplitudes $\mathcal{\tilde A}_{\Delta_1,\cdots,\Delta_n}(w_i,\bar w_i)$ are  \begin{align}
\boxed{~
\mathcal{\tilde A}_{\Delta_1,\cdots,\Delta_n}(w_i,\bar w_i) 
\equiv \prod_{i=1}^n \int_0^\infty {dy_i\over y_i^3}  \int dz_i d\bar z_i\,
G_{\Delta_i} (y_i,z_i, \bar z_i; w_i ,\bar w_i)\,\mathcal{A}(m_j \hat p_j^\mu)\,,~}
\end{align}
where $\hat p_j^\mu \equiv \hat p^\mu (y_j,z_j,\bar z_j)$  is given by \eqref{map} satisfying $\hat p_i^2=-1$.  By construction $\mathcal{\tilde A}_{\Delta_1,\cdots,\Delta_n}(w_i,\bar w_i)$  transforms covariantly under $SL(2,\mathbb{C})$,
\begin{align}
\mathcal{\tilde A}_{\Delta_1,\cdots,\Delta_n}\left({ aw_i +b\over cw_i+d},{\bar a\bar w_i+\bar b\over \bar c\bar w_i+\bar d}\right) =\left( \prod_{i=1}^n |c w_i +d|^{2\Delta_i} \right) \mathcal{\tilde A}_{\Delta_1,\cdots,\Delta_n}(w_i,\bar w_i)\,.
\end{align}

\subsection{Analytic Continuation and the Massless Wavefunction}
The integral expression \eqref{integral} is only a formal definition for the conformal primary wavefunction since the integral is divergent for real mass $m$.  More rigorously, we should define our conformal primary wavefunction by analytic continuation of the integral expression \eqref{integral} from an unphysical region.  When the mass is  purely imaginary $m\in -i \mathbb{R}_+$ and $X^\mu$ lies inside the future lightcone, the outgoing wavefunction \eqref{integral} is convergent and can be evaluated as
\begin{align}\label{prebessel}
&\phi^+_{\Delta,m}( X^\mu;w,\bar w)  = {4\pi\over |m|} {(\sqrt{-X^2} )^{\Delta-1}\over (-X^\mu q_\mu)^\Delta} K_{\Delta-1} \left( |m|\sqrt{-X^2}\right)\,,\notag\\
&\text{if}~~~~~~X^0>0\,,~~~~X^\mu X_\mu <0\,,~~~~~~m\in -i\mathbb{R}_+\,,
\end{align}\label{null}
where $q^\mu$ is a null vector in $\mathbb{R}^{1,3}$ defined as
\begin{align}
q^\mu = \left( 1+|w|^2 , w+\bar w, -i (w-\bar w) , 1-|w|^2\right)\,.
\end{align}
After landing on the Bessel function expression \eqref{prebessel}, we can then analytically continue it to real mass $m$ and other regions in $\mathbb{R}^{1,3}$, 
\be \label{asf}\phi^\pm_{\Delta,m}( X^\mu;w,\bar w)  = {4\pi\over im} {(\sqrt{-X^2} )^{\Delta-1}\over (-X^\mu q_\mu \mp i \epsilon)^\Delta} K_{\Delta-1} \left( im\sqrt{-X^2} \right) \,.\ee
We have introduced an $i\epsilon$ prescription to regularize the integral \eqref{integral} in the case of real mass $m$. 
In practice, the integral representation \eqref{integral} will however prove to be more convenient to compute the scattering amplitudes of these conformal primary wavefunctions.

We note that at late times inside the future lightcone, the wave equation takes the asymptotic form 
\be (\p^\mu\p_\mu -m^2)\phi=(-\p_\tau^2-{3 \over \tau}\p_\tau-m^2+\cdots)\phi\,,~~~~~\tau^2=-X^\mu X_\mu \to \infty\,.\ee
This is solved to leading nontrivial order at large $\tau$ by 
\be 
\phi={e^{\pm im\tau}\over \tau^{3/2}}\phi^{(0)}(y,z,\bz)+\cdots\,,
\ee
where $\phi^{(0)}$ is $any$ function on $H_3$.  One may check that (\ref{asf}) with $X^2<0$ takes this form. On the other hand, outside the lightcone near spatial infinity we have 
\be 
(\p^\mu\p_\mu -m^2)\phi=(\p_\sigma^2+{3 \over \sigma}\p_\sigma-m^2+\cdots)\phi\,,~~~~~\sigma^2=X^\mu X_\mu \to \infty\,.
\ee
This is solved to leading nontrivial order at large $\sigma$ by 
\be \phi={e^{\pm m\sigma}\over \sigma^{3/2}}\tilde \phi^{(0)}(y,z,\bz)+\cdots,\ee
with $\tilde \phi^{(0)}$ any function on dS$_3$. One may verify that (\ref{asf}) with  $X^2>0$ takes this form.\footnote{One should take the square root in \eqref{asf} corresponding to the decaying exponent when $X^\mu$ is outside the lightcone.}


From the Bessel function expression \eqref{asf} we can take the $m\to 0$ limit to obtain the massless conformal primary wavefunction  (assuming $\text{Re}(\Delta)>1$),\footnote{Here we have dropped an overall constant factor compared to the massless limit of \eqref{asf}.}
\begin{align}
\phi^\pm_{\Delta,m=0} (X^\mu;w,\bar w) = {1\over (-X^\mu q_\mu \mp i \epsilon)^\Delta}\,.
\end{align}
The massless conformal primary wavefunction has been considered in \cite{deBoer:2003vf,Campiglia:2015qka,Campiglia:2015kxa,Campiglia:2015lxa,Cheung:2016iub}.

\section{Massive Three-Point Amplitude}\label{sec:3pt}
In this section we will consider the tree-level three-point amplitude $\mathcal{ \tilde A}(w_i,\bar w_i)$ of the conformal primary wavefunction \eqref{integral} $\phi^\pm_{\Delta_i,m_i } ( X^\mu;w_i,\bar w_i)$, interacting through a local cubic vertex in $\mathbb{R}^{1,3}$,
\be
  \mathcal{L} \sim {\lambda } \phi_1 \phi_2  \phi_3+\cdots\,. 
\ee
The  three point scattering amplitude for plane waves is then simply 
\be 
\mathcal{A}(p_i)= i(2\pi)^4\lambda\, \delta^{(4)}(- p_1+p_2+p_3)\,.
\ee
For conformal primary wavefunctions we have 
\begin{align}\label{iop}
\mathcal{ \tilde A}(w_i,\bar w_i)&  = i \lambda \int d^4 X\,   \phi^-_{\Delta_1,m_1} (X^\mu;w_1,\bar w_1)
  \prod_{i=2}^3\phi^+_{\Delta_i ,m_i} (X^\mu;w_i,\bar w_i)\,,
\end{align}
where we take the first particle to be incoming and the other two be outgoing.  
The three-point amplitude is fixed by the $SL(2,\mathbb{C})$ covariance \eqref{highestweight} to be proportional to the standard three-point function in a two-dimensional CFT:
\be \mathcal{ \tilde A}(w_i,\bar w_i)\sim {\lambda \over |w_1-w_2|^{\Delta_1+\Delta_2-\Delta_3}|w_2-w_3|^{\Delta_2+\Delta_3-\Delta_1}|w_3-w_1|^{\Delta_3+\Delta_1-\Delta_2}}\,,\ee
but it is nevertheless satisfying to see this formula explicitly appear in a 4D scattering amplitude. 
We wish to determine the finite proportionality constant which is a function of the masses $m_i$ and the conformal dimensions $\Delta_i$. In general these are given by the integral expression (\ref{iop}) which may not be possible to analytically evaluate.  We will compute this  constant explicitly in the near-extremal case when the mass  of the first particle is slightly heavier than the sum of those of the other two.    In this case  the integral simplifies considerably and the three-point amplitude reduces to the tree-level three-point Witten diagram in $H_3$.

Let the mass of the first particle be $2(1+\e) m$ with $\e\ge 0$ and the masses of the other two particles be $m$.  Evaluating the $X^\mu$-integral, we arrive at the following expression for the scalar three-point amplitude,\footnote{We will use the integral representation \eqref{integral} of the conformal primary wavefunction to simplify the calculation of the amplitude and eventually make contact with the Witten diagram in $H_3$ in the near extremal limit. However, as discussed at the end of Section \ref{sec:wavefunction}, the integral representation  of our the conformal primary wavefunction is divergent for real mass $m$ and the proper definition requires an analytic continuation from the Bessel function expression \eqref{prebessel}. Nonetheless, we will see that the three-point amplitude computed using the integral representation  turns out to be finite.} 
\begin{align}
\mathcal{ \tilde A}(w_i,\bar w_i) =i(2\pi)^4 \lambda m^{-4}
\left( \prod_{i=1}^3 \int_0^\infty   {dy_i \over y_i^3} \int dz_id\bar z_i \, \right)
\prod_{i=1}^3 G_{\Delta_i} (y_i , z_i,\bar z_i ;  w_i,\bar w_i)\,
\delta^{(4)}(-2(1+\e) \hat p_1 +\hat p_2+\hat p_3)\,,
\end{align}
where $\hat p_i^\mu \equiv  \hat p^\mu (y_i , z_i,\bar z_i)$ as defined in \eqref{map}.  Note the integral does \textit{not} take the form of a tree-level three-point Witten diagram in $H_3$ for general $\e\ge0$.

 We now perform the $y_3,z_3,\bar z_3$-integrals to get rid of three delta functions. As a result, we have
\begin{align}
&\mathcal{ \tilde A}(w_i,\bar w_i)  =i 2(2\pi)^4\lambda m^{-4}\left(\int_0^\infty   {dy_1\over y_1^3} \int dz_1d\bar z_1\right)
\left(\int_0^\infty   {dy_2\over y_2^{3}} \int dz_2d\bar z_2\right)\prod_{i=1}^3 G_{\Delta_i} (y_i , z_i,\bar z_i ; w_i,\bar w_i)\notag\\
&\times
{1\over 2(1+\e) y_1^{-1} -y_2^{-1}}  \, 
\delta\left({2(1+\e)\over y_1-2(1+\e) y_2}\left[ -2\e y_1y_2+ (y_2-y_1)^2  +|z_2-z_1|^2\right]\right)\,,
\end{align}
where we have replaced
\begin{align}\label{y3}
&y_3 = {1\over 2(1+\e) y_1^{-1} -y_2^{-1}}\,,~~~~~~~~
z_3  = {2(1+\e)y_1^{-1} z_1 -y_2^{-1}  z_2 \over 2(1+\e)y_1^{-1} -y_2^{-1}}\,.
\end{align}
The overall factor 2 is due to the Jacobian coming from  rearranging the arguments of the delta functions. 
Now let us perform a change of variables from $(y_2,z_2,\bar z_2)$ to $(R,\theta, \phi)$,
\begin{align}\label{y2}
&y_2= y_1 + R\cos\theta\,,~~~~~~
z_2= z_1 + R \sin\theta \, e^{i\phi}\,,~~~~~~~0\le \theta \le \theta_*(y_1,R)\,.
\end{align}
The upper bound of $\theta$ is given by
\begin{align}
\theta_*(y_1,R)=
\begin{cases}
\pi\,, ~~~~~~~~~~~~~~~~~~~~~\text{if}~~~R<y_1\,,\\
\cos^{-1} \left( -{y_1\over R}\right)\,,~~~~~~~\text{if}~~~R\ge y_1\,.
\end{cases}
\end{align}
which comes from the constraint $\cos\theta = {y_2 -y_1 \over R } \ge - {y_1\over R}$.  
We can then rewrite the three-point amplitude as
\begin{align}
\mathcal{ \tilde A}(w_i,\bar w_i)  &= i 2(2\pi)^4\lambda m^{-4} \int_0^\infty   {dy_1\over y_1^{3}} \int dz_1d\bar z_1
\int_0^\infty  dR R^2  \int_0^{\theta_*(y_1,R)} d\theta \sin\theta \int_0^{2\pi} d\phi 
 \notag\\
&\times \, 
\prod_{i=1}^3 G_{\Delta_i} (y_i , z_i,\bar z_i ;w_i,\bar w_i)
 {y_1 \over \left(  ( 2\e+1) y_1  +2(1+\e) R\cos\theta\right) 
 \left( y_1 +R\cos\theta \right)^2
 } \, 
 \notag\\
 &
\times\delta\left({2(1+\e)\over 
( 2\e+1) y_1  +2(1+\e) R\cos\theta
}\left[ R^2 -2\e y_1(y_1+R\cos\theta)\right]\right)\,,
\end{align}
with $y_2,z_2,\bar z_2$ and $y_3, z_3,\bar z_3$ replaced by \eqref{y2} and \eqref{y3}.
The delta function has support on
\begin{align}
R= \sqrt{\e}  \,\sqrt{2 +\e\cos^2\theta}\, y_1 +\e y_1\cos\theta \,.
\end{align}

So far we have not taken any limit on the masses $2(1+\e) m,m,m$ of the three particles. Now let us consider the near extremal limit $\e\to0$. In this limit the three  momenta $m_i\hat p_i$ become collinear and the corresponding points $(y_i,z_i,\bar z_i)$ in $H_3$ become coincident.  To leading order in $\sqrt{\e}$, the solution of $R$ can be approximated by
\begin{align}
R = \sqrt{2\e} y_1  +\mathcal{O}(\e)\,.
\end{align}
In the near extremal limit we have $R\sim 0$, hence the three bulk points $y_i,z_i,\bar z_i$ in $H_{3}$ become coincident as commented above. We can then bring the three bulk-to-boundary propagators outside the $R,\theta,\phi$-integral.  Next, by a simple power counting of $R$, we find that the three-point amplitude of the conformal primary wavefunction is zero when $\e=0$, which is related to the fact that the the phase space vanishes for marginal decay process. We should proceed to the  subleading order in the near extremal expansion to obtain a nonzero answer, which is\footnote{Note that in the near extremal limit $\e\to0$, the upper bound $\theta_*(y_1,R)$ of the angular coordinate becomes $\pi$ on the support of the delta function.}
\be\label{main}
&\mathcal{ \tilde A}(w_i,\bar w_i)  ={ i (2\pi)^5 \lambda \over m^{4}} \int_0^\infty   {dy_1\over y_1^{3}} \int dz_1d\bar z_1
\prod_{i=1}^3 G_{\Delta_i} (y_1 , z_1,\bar z_1 ;w_i,\bar w_i)
 \notag\\
&~~~~~~~~~~~~\times {1 \over y_1} \int_0^\infty  dR R^2  \int_0^{\pi} d\theta \sin\theta \, 
 \, \delta\left( R^2 -2\e y_1^2\right) +\mathcal{O}(\e)\,\\
 & = { i2^{11\over2}\pi^5 \lambda\over m^4} \sqrt{\e} \left( \int_0^\infty   {dy_1\over y_1^3} \int dz _1d\bar z_1
\prod_{i=1}^3 G_{\Delta_i} (y_1 , z_1,\bar  z_1 ;w_i,\bar w_i)\right) +\mathcal{O}(\e)\, \notag\\&={ i 2^{9\over2}\pi^6\lambda \Gamma({\Delta_1+\Delta_2+\Delta_3-2\over 2}) \Gamma({\Delta_1+\Delta_2-\Delta_3\over 2})\Gamma({\Delta_1-\Delta_2+\Delta_3\over 2})\Gamma({-\Delta_1+\Delta_2+\Delta_3\over 2})\sqrt{\e}\over m^4
\Gamma(\Delta_1)\Gamma(\Delta_2)\Gamma(\Delta_3) |w_1-w_2|^{\Delta_1+\Delta_2-\Delta_3}|w_2-w_3|^{\Delta_2+\Delta_3-\Delta_1}|w_3-w_1|^{\Delta_3+\Delta_1-\Delta_2}} +\mathcal{O}(\e)\,,\notag
\ee
where the term in the parentheses in the second to last line is precisely the tree-level three-point Witten diagram in $H_3$, which was evaluated in \cite{Freedman:1998tz}.    
Hence the near extremal massive three-point amplitude takes the form of the  three-point function of scalar primaries with conformal dimensions $\Delta_i$ in a two-dimensional CFT.

\section*{Acknowledgements}
We are  grateful to  T. Dumitrescu, P. Mitra, M. Pate, B. Schwab, D. Simmons-Duffin, and A. Zhiboedov for useful conversations.   This work was supported in part by NSF grant 1205550.  S.P. is supported by the NSF and by the Hertz Foundation through a Harold and Ruth Newman Fellowship.   S.H.S. is supported by the National
Science Foundation grant PHY-1606531.

\appendix
\section{Klein-Gordon Inner Product}\label{sec:KG}
In this section we evaluate the Klein-Gordon inner product between two conformal primary solutions with the same mass $m$ and generic complex weights  $\Delta_{1,2}$.  $SL(2,\mathbb{C})$ implies this must vanish for $\Delta_1\neq \Delta_2^*$, while some kind of delta function is expected at $\Delta_1=\Delta_2^*$. 

The  Klein-Gordon inner product between two outgoing wavefunctions $\phi^+_{\Delta_1 , m } (X^\mu; w_1,\bar w_1)$ and  $\phi^+_{\Delta_2 , m } (X^\mu; w_2,\bar w_2)$ evaluated on the slice $X^0=0$ is  \begin{align}{\label{eq:ipt}}
( \phi^+_1 , \phi^+_2)&=- i \int d^3 \vec X~\left[\,  
\phi^+_{\Delta_1,m} ( X^\mu; w_1,\bar w_1) \partial_{X^0}\phi_{\Delta_2,m}^{+*} ( X^\mu; w_2,\bar w_2)\right.\notag\\
&\left.
~~~~~~~~~~~~~~~~
-\partial_{X^0}\phi^+_{\Delta_1,m} ( X^\mu; w_1, \bar w_1) \phi_{\Delta_2,m}^{+*} ( X^\mu; w_2,\bar w_2)
\right]
\notag\\
=&(2\pi)^3m^{-2}\left( \prod_{i=1}^2 \int_0^\infty {dy_i \over y_i^3} \int dz_i d\bar z_i\right) G_{\Delta_1}(y_1,z_1,\bar z_1;w_1,\bar w_1) G_{\Delta_2} ^*(y_2 ,z_2,\bar z_2;w_2 ,\bar w_2)\notag\\
\times& \left( {1+y_1^2 +|z_1|^2 \over 2y_1 }+{1+y_2^2 +|z_2|^2 \over 2y_2}\right)
\delta^{(2)}\left( {z_1\over y_1} -{z_2\over y_2}\right)
\delta\left({1-y_1^2 -|z_1|^2 \over 2y_1 }-{1-y_2^2 -|z_2|^2 \over 2y_2 }\right)\notag\\
=&2(2\pi)^3m^{-2} \int_0^\infty   {dy\over y^{3}} \int dzd\bar z
 G_{\Delta_1} (y , z,\bar z;w_1,\bar w_1)G^*_{\Delta_2} (y , z,\bar z ;w_2,\bar w_2)\,.
 \end{align}
  Using the Feynman trick,
\begin{align}
{1\over A^a B^b }  = {\Gamma(a+b) \over \Gamma(a) \Gamma(b) } \int_0^1 d\alpha { \alpha^{a-1} (1-\alpha)^{b-1} \over ( \alpha A + (1-\alpha)B)^{a+b} }\,,
\end{align}
we can perform the integrals in $y,z,\bar z$ to obtain
 \begin{align}
(\phi_1^+,\phi_2^+)=2(2\pi)^3 m^{-2}{\pi \Gamma({\Delta_1+\Delta_2^*-2\over2 })
\Gamma({\Delta_1+\Delta^*_2\over2 }) 
\over 2\Gamma(\Delta_1)\Gamma(\Delta_2^*)  |w_1-w_2|^{\Delta_1+\Delta_2^*}} \int_0^1 d\alpha \alpha^{{\Delta_1-\Delta_2^*\over2}-1 }(1-\alpha)^{{-\Delta_1 +\Delta^*_2\over2}-1} \,.
\end{align}
Here $ \Delta_2^*$ is the complex conjugate of $\Delta_2$. If we let $\eta={\Delta_1-{\Delta}^*_2\over2 }$,  $\alpha=\frac{e^u}{e^u+e^{-u}}$
\begin{align}
\int_0^1 d\alpha \alpha^{\eta-1 }(1-\alpha)^{-\eta-1}=2\int_{-\infty}^{\infty} du e^{2u\eta}\,.
\end{align}
This is divergent if $\eta$ is real, and equals to $2\pi \delta(\lambda)$ if $\eta=i\lambda$ is pure imaginary.
 Therefore, in order to have a  delta function normalizable inner product,  we require $\Delta_i$'s to be complex numbers with the same real part, $\Delta_1=a+i\lambda_1$, $\Delta_2=a+i\lambda_2$ ($a,\lambda_i\in \mathbb{R}$).  The Klein-Gordon inner product for complex conformal dimensions, equal mass conformal primary wavefunction is,
\begin{align}{\label{eq:ipt}}
(\phi_1^+,\phi_2^+)=64\pi^5 m^{-2} {1\over \left( \Delta_1+\Delta_2^*-2\right) |w_1-w_2|^{\Delta_1+\Delta_2^*}} \delta(\lambda_1+\lambda_2)\,.
\end{align}

\bibliography{massivefinal}{}
\bibliographystyle{utphys}

\end{document}